\documentclass{nature1}

\usepackage{epsfig}

\def\gtrsim{\mathrel{\hbox{\rlap{\hbox{\lower4pt\hbox{$\sim$}}}\hbox{$>$}}}}
\def\lesssim{\mathrel{\hbox{\rlap{\hbox{\lower4pt\hbox{$\sim$}}}\hbox{$<$}}}}

\def\degree{$^\circ$}
\def\arcs#1{$#1''$}
\def\arcsa#1#2{$#1^{''}{\textrm{.}}#2$}  % for science

\def\solarmass{$M_\odot$}
\def\solarmasse{M_\odot}

\def\mJyb{mJy beam$^{-1}$}

\def\cmc{cm$^{-3}$}

\def\degr{\hbox{$^\circ$}}
\def\micron{$\mu$m}

\def\ra#1#2#3#4{#1^\mathrm{h} #2^\mathrm{m} #3^\mathrm{s}_{^\textrm{.}} #4}
\def\dec#1#2#3#4{#1\degr #2' #3^{''}{\textrm{.}}#4}  % for science

\def\mH2{m_{\textrm{\scriptsize H}_2}}
\def\Ro{R_\textrm{\scriptsize 0}}

\def\To{T_\textrm{\scriptsize 0}}
\def\no{n_\textrm{\scriptsize 0}}

\def\H2{H$_2$}
\def\N2HP{N$_2$H$^+$}
\def\HCOP{HCO$^+$}

\def\NH3{NH$_3$}

\def\SOt{$N_J=8_9-7_8$}

\def\HCOP{HCO$^+$}

\title{First Detection of Equatorial Dark Dust Lane in a Protostellar Disk
at Submillimeter Wavelength}

\author{Chin-Fei Lee,$^{1,2*}$ Zhi-Yun Li,$^{3}$ Paul. T.P Ho,$^{1,4}$ Naomi
Hirano,$^{1}$ Qizhou Zhang,$^{4}$ Hsien Shang$^{1}$} 

\begin{document}

\maketitle

\begin{affiliations}
 \item Academia Sinica Institute of Astronomy and Astrophysics, P.O. Box 23-141, Taipei 106, Taiwan
 \item Graduate Institute of Astronomy and Astrophysics, National Taiwan 
   University, No.  1, Sec.  4, Roosevelt Road, Taipei 10617, Taiwan
 \item Astronomy Department, University of Virginia, Charlottesville, VA, USA
 \item Harvard-Smithsonian Center for Astrophysics, 60 Garden Street, Cambridge, MA 02138
\end{affiliations}

% with a protostellar mass of $\sim$ 0.2-0.3
%\solarmass{}\cite{Lee2014,Codella2014} and an estimated age of only $\sim$
%40,000 years\cite{Lee2014}.

%With an unprecedented resolution of $\sim$ 8 AU (with 1 AU being the
%distance between the Earth and Sun) and sensitivity of ALMA,

\noindent {\bf (Teaser) Directly detecting and characterizing small disks
around the youngest protostars through high-resolution imaging with ALMA}\\

\begin{abstract}
In the earliest (so-called ``Class 0") phase of sunlike (low-mass) star
formation, circumstellar disks are expected to form, feeding the
protostars.  However, such disks are difficult to resolve spatially because
of their small sizes.  Moreover, there are theoretical difficulties in
producing such disks in the earliest phase, due to the retarding effects of
magnetic fields on the rotating, collapsing material (so-called ``magnetic
braking").  With the Atacama Large Millimeter/submillimeter Array (ALMA),
it becomes possible to uncover such disks and study them in detail.  HH 212
is a very young protostellar system.  With ALMA, we not only detect but also
spatially resolve its disk in dust emission at submillimeter
wavelength.  The disk is nearly edge-on and has a radius of $\sim$ 60
AU. Interestingly, it shows a prominent equatorial dark lane sandwiched
between two brighter features, due to relatively low temperature and high
optical depth near the disk midplane. For the first time, this dark lane is
seen at submillimeter wavelength, producing a ``hamburger"-shaped
appearance that is reminiscent of the scattered-light image of an edge-on
disk in optical and near infrared. Our observations open up an exciting
possibility of directly detecting and characterizing small disks around the
youngest protostars through high-resolution imaging with ALMA, which
provides strong constraints on theories of disk formation.

\end{abstract}

\clearpage

\subsection{INTRODUCTION}\mbox{}\\ Stars are formed inside molecular cloud
cores through gravitational collapse.  The details of the process, however,
are complicated by the presence of magnetic fields and rotation.  In theory,
(rotationally supported) circumstellar disks are expected to form
inside collapsing cores around protostars, feeding the protostars.  Such
disks have been detected with radii up to $\sim$ 500 AU in the late (i.e.,
Class II or T Tauri) phases of sunlike star
formation\cite{Simon2000,Perez2016}.  Such disks must have started in the
earliest (Class 0) phase, as claimed in a few Class 0 sources, e.g., HH
211\cite{Lee2009}, L 1527\cite{Tobin2012}, and recently VLA
1623\cite{Murillo2013}.  However, those disks, with radii $<$ 150 AU, have
not been well resolved spatially, especially in the vertical direction,
because of insufficient resolution.

In models of non-magnetized core collapse, a circumstellar disk can
indeed form as early as in the Class 0 phase\cite{Terebey1984}.  However, a
realistic model should include magnetic fields, because molecular cores are
found to be magnetized\cite{Chapman2013}.  In the presence of a dynamically
important magnetic field, when and how disks form becomes uncertain, because
the field can remove angular momentum of the collapsing material
efficiently, leading to the so-called ``magnetic braking catastrophe" in
disk formation\cite{Li2014}.  In those cases, a flattened envelope called
the pseudodisk can form around the central source, but not a large
rotationally supported disk\cite{Allen2003}.  A misalignment between the
magnetic field and the axis of rotation\cite{Joos2012,Li2013} and non-ideal
MHD effects\cite{Tomida2015} are sometimes able to solve this catastrophe,
but not always.  Even when a disk is formed, if a significant fraction
of the magnetic flux of the dense core is dragged into it, the disk would
rotate at a sub-Keplerian speed because of magnetic support\cite{Shu2008}.

%, with important consequences for its structure and dynamics
%(Lizano et al.  2016, ApJ, 817, 35).

HH 212 is a nearby protostellar system deeply embedded in a compact
molecular cloud core in the L1630 cloud of Orion at a distance of $\sim$ 400
pc.  The central source is the Class 0 protostar IRAS 05413-0104, with
a mass of $\sim$ 0.2-0.3 \solarmass{}\cite{Lee2014,Codella2014} and a
bolometric luminosity $L_\textrm{\scriptsize bol}\sim$ 9 $L_\odot$ (updated
for the distance of 400 pc)\cite{Zinnecker1992}.  It is very young with an
estimated age of only $\sim$ 40,000 yrs\cite{Lee2014}.  It drives a powerful
bipolar jet\cite{Zinnecker1998}.  Accordingly to current jet
models\cite{Konigl2000,Shu2000}, a circumstellar disk is required to
launch the jet.  Recent observations at a resolution of $\sim$ \arcsa{0}{5}
(200 AU) shows a flattened envelope and a tentative compact disk around the
central source in $\sim$ 850 \micron{} dust continuum\cite{Lee2014}.  The
gas kinematics in \HCOP{} shows that the flattened envelope is infalling
with small rotation (i.e., spiraling) into the central source, which is
expected for the pseudodisk in the models of magnetized core
collapse\cite{Allen2003}.  Inside this structure, the \HCOP{} and C$^{17}$O
kinematics indicate the presence of a rotationally supported
circumstellar disk around the central source, with a radius $\lesssim$ 100
AU\cite{Lee2014,Codella2014}.  Here we report ALMA observations in dust
emission at $\sim$ 850 \micron{} that clearly resolves the disk spatially.

\subsection{RESULTS}\mbox{}\\
Figure 1a shows the inner part of the jet within $\sim$ 12000 AU (\arcs{30})
of the source, where it was mapped in H$_2$ with the Very Large Telescope at
$\sim$ \arcsa{0}{35} resolution\cite{McCaughrean2002} and in SiO and CO with
ALMA at $\sim$ \arcsa{0}{5} resolution\cite{Lee2015}, in order to show the
physical relationship of the envelope/disk with the jet.  The jet is highly
collimated with a position angle (PA) of $\sim$ 23\degree{}.  In this
figure, a reflection nebula is also seen in continuum at 2.12 \micron{} in
the infrared near the source, with a dust lane roughly perpendicular to the
jet axis.  This dust lane is due to dust extinction of the extended envelope
detected in \NH3{}\cite{Wiseman2001}.  The jet was not detected in H$_2$
near the source due to the same dust extinction.  However, it was clearly
detected in SiO down to the central source (Figure 1b).

At the submillimeter wavelength, dust extinction from the envelope is
negligible, and thus we can detect the dust emission down to the small
scales at which the circumstellar disk is formed.  Previously, a
flattened envelope was detected in continuum at 850 \micron{} deep inside
the extended envelope within $\sim$ 1000 AU of the central source at $\sim$
\arcsa{0}{5} (200 AU) resolution\cite{Lee2014} (Figures 1a and 1b for zoom
in).  Zooming into the center at a higher resolution of $\sim$ \arcsa{0}{1}
(40 AU) (Figure 1c), we see the inner part of the envelope extending along
the major axis with PA $\sim$ 123\degree{} (as indicated by the white
lines), roughly perpendicular to the jet axis.  Zooming further into the
center at an unprecedented resolution of $\sim$ \arcsa{0}{02} (8 AU) (Figure
1d), we see the central peak being resolved into a bright disklike structure
with a major axis at PA $\sim$ 113\degree{} (as indicated by the white
lines), exactly perpendicular to the jet axis.  At this high resolution,
the noise level becomes $\sim$ 0.08 \mJyb{} (see Methods), therefore,
the envelope, with a flux density of $<$ 0.11 \mJyb{} (estimated from Figure
1c), can not be detected.  There is a big jump (by a factor of $\sim$ 10) in
the brightness temperature (hereafter $T_b$) from $\sim$ 3 K (or 3 \mJyb{}
in a \arcsa{0}{1} beam in Figure 1c) in the innermost envelope to $\sim$ 30
K (or 1.1 \mJyb{} in a \arcsa{0}{02} beam in Figure 1d) in the outer part of
the disklike structure, indicating that the disklike structure is
physically distinct from the envelope and thus can be naturally identified
as the circumstellar disk.  As discussed below, since the
emission in the outer disk is becoming optically thick, the density in the
outer disk is expected to be more than 10 times higher than that in the
innermost envelope. This jump could be due to the density jump produced by
an accretion shock in disk formation\cite{Krasnopolsky2002,Lee2014}.

Being perpendicular to the jet, the disk must be nearly edge-on with the
nearside tilted to the southwest, because the jet has a small inclination
angle of $\sim$ 4$\pm2$\degree{} to the plane of the
sky\cite{Claussen1998,Lee2007}, with the northeastern component tilted
slightly toward us.  A prominent dark lane is seen running along the major
axis of the disk.  The thickness of the dark lane increases with the
increasing distance from the center, giving the impression that the
disk is flared, although detailed modeling is required to ascertain whether
this is indeed the case. A cut along the dark lane shows that  the flux
density increases rapidly from $\sim$ 0.4 \mJyb{} ($T_b\sim 11$ K, 5
$\sigma$) at $\sim$ 68 AU (\arcsa{0}{17}) to $\sim$ 1.8 \mJyb{} ($T_b\sim
50$ K) at $\sim$ 40 AU (\arcsa{0}{1}), and then slowly increases to $\sim$
2.1 \mJyb{} ($T_b\sim 58$ K) at the center (see Figure 2), suggesting that
the emission is optically thick along the dark lane, except near the edge.
This suggests that mm-size particles may dominate the emission.
The half-width at half maximum of the intensity profile is $\sim$ 60 AU
(\arcsa{0}{15}).  The dark lane is sandwiched by two brighter regions, one
above at $\sim$ 3.3 \mJyb{} ($T_b\sim 90$ K) in the northeast and one below
at $\sim$ 2.9 \mJyb{} ($T_b\sim 80$ K) in the southwest. The two regions
are from the surface of the disk.  They are brighter because the disk
surface is expected to be warmer than the midplane, due to the heating by
the radiation of the central protostar\cite{Chiang1997} and possible
interaction of the disk surface with the wind from the innermost
disk\cite{Li1996}.  Since the nearside of the disk is tilted slightly to the
southwest, the emission in the northeast is less absorbed by the nearside of
the disk and thus brighter than the emission in the southwest.  

\subsection{DISCUSSION}\mbox{}\\
Previous observations of the disk emission at 1.4 mm and 0.85 mm
suggested a spectral index of $\sim$ 2.6\cite{Lee2008}, and thus a dust
opacity spectral index $\beta \sim 0.6$,
 similar to the values often obtained for disks in more evolved (Class
II) objects\cite{Natta2007}.
This $\beta$ indicates that grain growth to mm size or larger\cite{Draine2006} has
started early in the formation of the disk, or perhaps even before this
phase, and that dust scattering is expected to contribute significantly
to the disk emission in submillimeter wavelengths\cite{Kataoka2015}.
Dust scattering should produce continuum polarization that, if detected, can
provide an independent probe of the grain properties, especially their sizes
\cite{Kataoka2015}. The total flux of the disk emission
is $\sim$ 157 mJy.  If the emission is optically thin and all thermal at a
temperature of 50~K, then the disk mass is $\sim 0.014$~M$_\odot$ for a dust
opacity of 0.054 cm$^2$ g$^{-1}$ at 850~$\mu$m\cite{Beckwith1990}.  This
estimate is quite uncertain, however, due to the large uncertainty in dust
opacity and potential effects of dust scattering and because the emission in
the outer disk is becoming optically thick.

%In the optically thin limit, the detected flux corresponds to a disk mass of
%$\sim 0.014$~M$_\odot$ for a dust opacity of 0.054 cm$^2$ g$^{-1}$ at
%850~$\mu$m (cite reference) and a temperature of 50~K.

To illustrate more quantitatively how the most important feature
observed in HH 212, a dark lane sandwiched by two brighter regions (i.e.,
the ``hamburger" appearance), may arise physically, we constructed a simple
toy model for disk emission in submillimeter.  The model is described in
the supplementary materials.  Briefly, the distributions of the density and
temperature of the disk are prescribed, with the surface warmer than the
midplane at a given radius.  Figure~3 shows the emission at 850 $\mu$m for
the model disk.  It broadly reproduces the major features observed in the HH
212 disk, including the dark lane, the bright regions above and below the
lane, with the emission above brighter than that below.  In this particular
example, the dark lane comes from the cooler outer part of the disk that
becomes optically thick at a radius around 40~AU (\arcsa{0}{1}) and the
brighter emission above and below the lane is due to the warmer surfaces of
the inner disk, in agreement with expectation.  We refrain from drawing
quantitative conclusions about the disk parameters from the model , except
to note that the radius of the model disk (68~AU) is not far from that
estimated based on the intensity profile in Figure 2 ($\sim 60$ AU) and that
the disk mass ($0.05$~M$_\odot$) is also consistent with the minimum mass
estimated under the optically thin assumption ($0.014$~M$_\odot$).  To do
better, major refinements would be needed, including a self-consistent
determination of the density and temperature structures and a treatment of
dust scattering, which is expected to be important for large grains.

%In the model, the disk
%has a radius of $\sim$ 68$\pm$12 AU (\arcsa{0}{17}$\pm$\arcsa{0}{03}), a
%scale height of $\sim$ 12$\pm$4 AU (\arcsa{0}{03}$\pm$\arcsa{0}{01}) at 36
%AU, and a mass of $\sim$ {\bf 0.05} \solarmass{}. 

Our observations show the first equatorial dark dust lane in a
protostellar disk in the earliest phase of star formation.  The detection
is made possible by the unprecedented resolution achieved by ALMA long
baseline.  The prominent dark lane and the large brightness
contrast with the envelope  are strongly suggestive of the disk-like
structure being a rotationally supported disk.  Indeed, there is already
strong indication for a Keplerian disk inside 100~AU of the protostar from
molecular line observations\cite{Lee2014,Codella2014}.  During the earliest
phase of star formation, the disks are expected to be rather small and
possibly massive, which make them likely optically thick and the velocity
profile difficult to measure, at least in the relatively short-wavelength
ALMA bands that are needed to resolve the (small) disks. 
In this case, the dust continuum may be the only viable way to detect and characterize these
youngest protostellar disks.  Our resolved observations of HH 212 disk
demonstrated that this is possible, at least for nearly edge-on systems.  It
raises the exciting possibility of detecting even smaller disks around
deeply embedded protostars. [Note
that our results of the HH 212 self-obscured disk could have 
general application to other Class 0 sources \cite{Li2017}].
 If small disks of tens of AUs (or smaller) turn
out to be common around the earliest protostars, it would imply that the
magnetic braking catastrophe in the theoretical literature of disk formation
is averted early on, plausibly through non-ideal MHD effects, which are
expected to be most efficient at decoupling the magnetic field from the bulk
material close to the central star where the density is the
highest\cite{Nakano2002}. 
Just as importantly, our observations open a window on the vertical
structure of the disks around deeply embedded protostars in the earliest
(Class 0) phase, which could ultimately be compared to those surrounding
more evolved young stars (Class I sources and T Tauri stars), potentially
yielding key insights on the processes of grain growth and settling that are
important to planet formation.

\subsection{MATERIALS AND METHODS}\mbox{}\\
Observations of the HH 212 protostellar system were carried out with
ALMA in Band 7 at $\sim$ 350 GHz in Cycles 1 and 3, with 32-45 antennas and
projected baselines ranging from $\sim$ 15 to 16200 m.  The total time on
the HH 212 system is $\sim$ 148 minutes.  We also introduce a parametrized
model for the disk emission in HH 212, in order to illustrate the formation
of the dark lane sandwiched between two brighter regions.  Details of our
data reduction and radiative transfer modeling are provided in Supplementary
Materials.

\subsection{SUPPLEMENTARY MATERIALS}\mbox{}\\
\noindent
Materials and Methods\\
Table S1. Observation Logs\\
Table S2. Correlator Setup for Cycle 1 Project\\
Table S3. Correlator Setup for Cycle 3 Project\\
Table S4. Calibrators and Their Flux Densities\\
References (34-36)
\clearpage
\newcommand\aap{{\it Astron. Astrophys.}}%M
\newcommand\apjl{{\it Astrophys. J. Lett.}}%M
\newcommand\apj{{\it Astrophys. J.}}%M
\newcommand\apjs{{ApJS}}%M
\newcommand\aj{{\it Astron. J.}}%M
\newcommand\araa{{ARAA}}%M
\newcommand\nat{{\it Nature}}%M
\newcommand\sci{{\it Science}}%M
\newcommand\mnras{{MNRAS}}%M

\subsection{REFERENCES AND NOTES}

\begin{addendum}

\item This paper makes use of the following ALMA data:
ADS/JAO.ALMA\#2012.1.00122.S and 2015.1.00024.S.  ALMA is a partnership of
ESO (representing its member states), NSF (USA) and NINS (Japan), together
with NRC (Canada), NSC and ASIAA (Taiwan), and KASI (Republic of Korea), in
cooperation with the Republic of Chile.  The Joint ALMA Observatory is
operated by ESO, AUI/NRAO and NAOJ.  {\bf Funding:} C.-F.L.  acknowledges
grants from the Ministry of Science and Technology of Taiwan (MoST
104-2119-M-001-015-MY3) and the Academia Sinica (Career Development Award). 
Z.-Y.L.  is supported in part by NSF AST1313083 and NASA NNX14AB38G, NH by
MoST 105-2112-M-001-026.  {\bf Author Contributions:} C.-F.L.  led the
project, analysis, discussion, and drafted the manuscript.  Z.-Y.L. 
participated in the analysis and discussion, and commented on the
manuscript.  All other coauthors contribute to scientific discussion.  {\bf
Competing Interests:} The authors declare that they have no competing
interests.  {\bf Data and materials availability:} All data needed to
evaluate the conclusions in the paper are present in the paper and/or the
Supplementary Materials.  Our ALMA data can also be obtained from the ALMA
Science Data Archive, https://almascience.nao.ac.jp/alma-data".  Additional
data available from authors upon request.

\end{addendum}
  
\clearpage
\def\putfig#1#2#3{\epsfig{scale=#1,angle=#2,figure=#3}}
%{\bf \noindent Figure Caption}
 
\newcounter{mfigure}[section]  
\newenvironment{mfigure}[1][]{\refstepcounter{mfigure}\par\medskip
   \noindent \textbf{Figure~\themfigure. #1} \rmfamily}{\medskip}

\begin{figure}
\centering
\putfig{0.9}{270}{f1.eps} %RGB_jet_cont.eps}
\end{figure}
\begin{mfigure}
%\caption
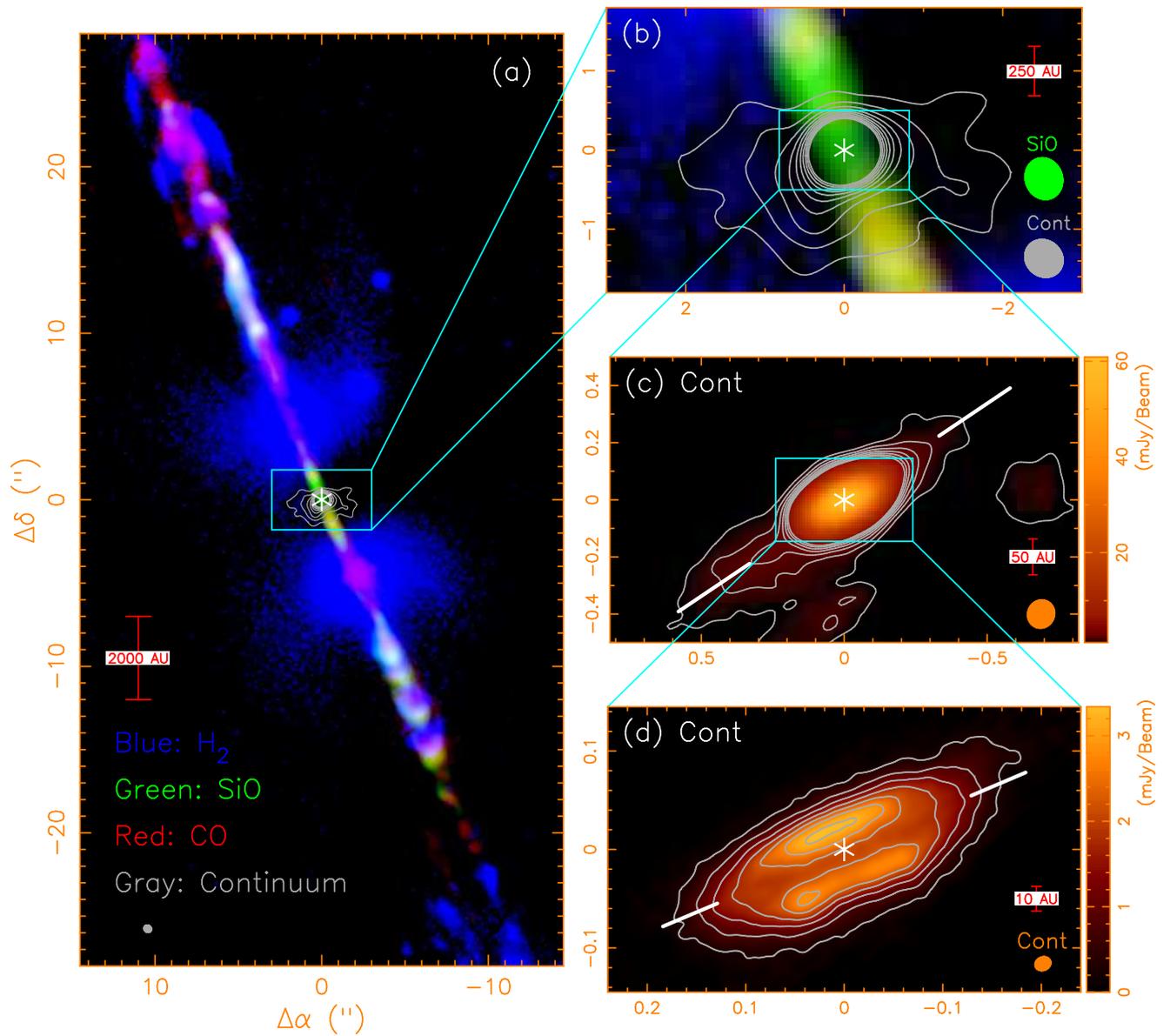
{(a) A composite image for the inner part of the HH 212 jet: blue shows
the map of H$_2$ $+$ 2.12 \micron{} continuum, obtained with the Very Large Telescope\cite{McCaughrean2002},
green and red show the SiO and CO maps, respectively, obtained with ALMA\cite{Lee2015}.
Gray contours show the continuum map of the envelope/disk at 850 \micron{} obtained with ALMA at $\sim$ \arcsa{0}{5}
resolution\cite{Lee2014}.
%Contours start at 0.135 K with a step of 0.162 K.
Contours start at 3.125 \mJyb{} with a step of 3.75 \mJyb{}.
(b) A zoom-in to the center for the jet and envelope/disk.
(c) A zoom-in to the center of the continuum at $\sim$ \arcsa{0}{1} resolution.
Contours start at 1.23 \mJyb{} with a step of 0.62 \mJyb{}.
(d) A zoom-in to the center of the continuum at $\sim$ \arcsa{0}{02} resolution.
Contours start at 0.29 \mJyb{} with a step of 0.49 \mJyb{}.
Asterisks mark the possible source position at
$\alpha_{(2000)}=\ra{05}{43}{51}{4086}$,
$\delta_{(2000)}=\dec{-01}{02}{53}{147}$, obtained by comparing to the model in 
Figure \ref{fig:contmod}. 
\label{fig:jet_cont}}
\end{mfigure}

\begin{figure}
\centering
\putfig{0.9}{270}{f2.eps} % cont_cut.eps}
\end{figure}
\begin{mfigure}
%\caption
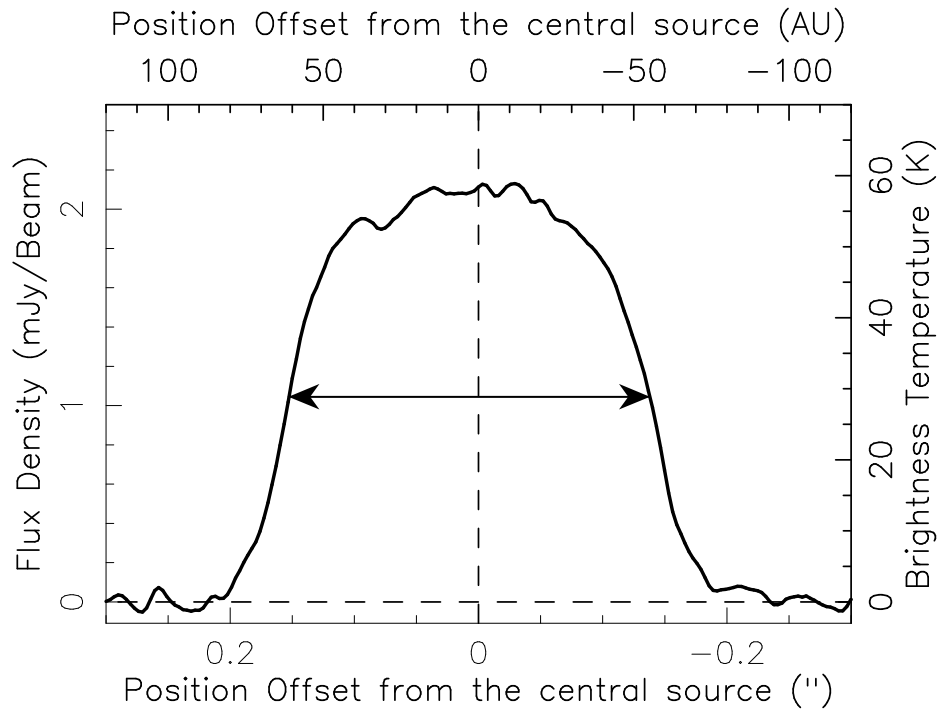
{Flux density (and corresponding brightness temperature)
of the continuum cut along the dark lane.
The vertical dashed line indicates the source position.
The double sized arrow marks the full-width at half maximum of the intensity
profile.
\label{fig:contcut}}
\end{mfigure}

\begin{figure}
\centering
%\putfig{0.65}{270}{f3a.eps} % 3d_disk.eps}
\putfig{0.7}{270}{f3.eps} % 3d_disk.eps}
\end{figure}
\begin{mfigure}
%\caption
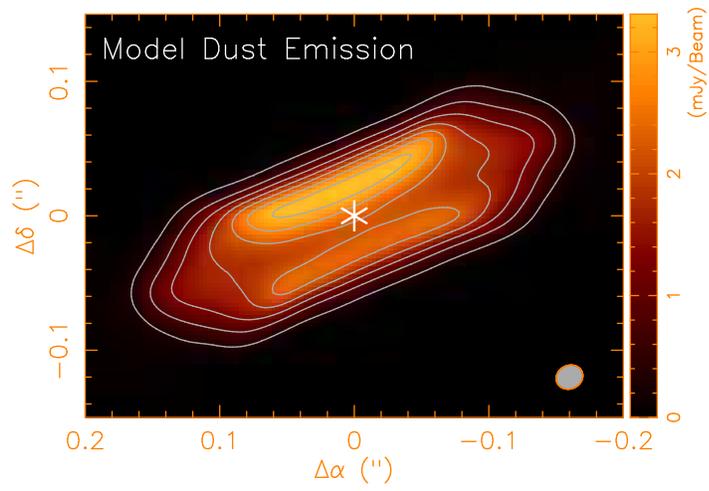
{Simulated observed continuum emission at 850 \micron{} derived from the model.
Contour levels are the same as in Figure \ref{fig:jet_cont}d.
\label{fig:contmod}}
\end{mfigure}

\clearpage

\noindent
{\bf \Huge Supplementary Materials}

\newcommand{\beginsupplement}{%
        \setcounter{table}{0}
        \renewcommand{\thetable}{S\arabic{table}}%
        \setcounter{figure}{0}
        \renewcommand{\thefigure}{S\arabic{figure}}%
     }

\beginsupplement

%\begin{methods}
\subsection{Materials and Methods}
\subsection{Observations} 

Observations of the HH 212 protostellar system were carried out with ALMA in
Band 7 at $\sim$ 350 GHz in Cycles 1 and 3, with 32-45 antennas and
projected baselines ranging from $\sim$ 15 to 16200 m (see Table
\ref{tab:obs}).  The Cycle 1 project was carried out with 2 executions, both
on 2015 August 29 during the Early Science Cycle 2 phase.  The Cycle 3
project was carried out with 2 executions in 2015, one on November 5 and one
on December 3, during the Early Science Cycle 3 phase.  For the Cycle 1
project, the correlator was set up to have 4 spectral windows, with one for
CO $J=3-2$ at 345.795991 GHz, one for SiO $J=8-7$ at 347.330631 GHz, one for
HCO$^+$ $J=4-3$ at 356.734288 GHz, and one for the continuum at 358 GHz (see
Table \ref{tab:corr1}).  For the Cycle 3 project, the correlator was more
flexible and thus was set up to include 2 more spectral windows, with one
for SO \SOt{} at 346.528481 GHz and one for H$^{13}$CO$^+$ $J=4-3$ at
346.998338 GHz (see Table \ref{tab:corr3}).  The total time on the HH 212
system is $\sim$ 148 minutes.

In this paper, we only present the observational results in continuum, which
traces the envelope and disk around the central source.  Line-free channels
were extracted from all the spectral windows to make the continuum with a
frequency centered at $\sim$ 352 GHz (or $\sim$ 850 \micron).  The data were
calibrated with the Common Astronomy Software Applications (CASA) package,
with quasars as bandpass, flux, and phase calibrators, as listed in Table
\ref{tab:calib} with their flux densities.  Two different weightings were
used to generate the continuum maps at two different resolutions.  One used
a robust factor of 2 (natural weighting) with the baselines from $\sim$ 15
to 2000 m in order to map the innermost envelope around the disk with enough
sensitivity and resolution.  This generates a synthesized beam (resolution)
of \arcsa{0}{100}$\times$\arcsa{0}{094} at a position angle (P.A.) of
$-$25\degree{} and a noise level of $\sim$ 0.42 \mJyb{} (0.44 K) (Figure 1c).  Another one
used a robust factor of 0.5 (super-uniform weighting) with all the
baselines, in order to resolve the disk structure.  In addition, data from
all baselines were included so that only structure with a size scale greater
than $\sim$ \arcs{2} could be filtered out in our image.  This generates a
synthesized beam of \arcsa{0}{0203}$\times$\arcsa{0}{0175} at a P.A.  of
$-$65\degree{} and a noise level of $\sim$ 0.08 \mJyb{} (2.2 K) (Figure 1d).

\begin{table}
%\scriptsize
\small
\centering
\caption{Observation Logs}
\label{tab:obs}
\begin{tabular}{llllll}
\hline
Cycle & Date & Array &Number of &Time on target & Baselines  \\
      & (YYYY-MM-DD) & Configuration & Antennas&(minutes) & (meter) \\
\hline\hline
1 & 2015-08-29 & C32-6   & 37 & 30  & 15$-$1466   \\ 
1 & 2015-08-29 & C32-6   & 35 & 30  & 15$-$1466   \\ 
3 & 2015-11-05 & C36-7/8 & 45 & 44  & 78$-$16196 \\ 
3 & 2015-12-03 & C36-7/8 & 32 & 44  & 17$-$6344  \\ 
\hline
\end{tabular}
\end{table}

\begin{table}
\small
\centering
\caption{Correlator Setup for Cycle 1 Project}
\label{tab:corr1}
\begin{tabular}{llllll}
\hline
Spectral  & Line or   & Number of & Central Frequency & Bandwidth & Channel Width\\
Window & Continuum & Channels  & (GHz)             & (MHz)     & (kHz) \\
\hline\hline
0 & CO J=3-2      & 3840  & 345.803 & 468.750  & 122.070  \\
1 & SiO J=8-7     & 3840  & 347.338 & 468.750  & 122.070  \\
2 & HCO$^+$ J=4-3 & 3840  & 356.742 & 468.750  & 122.070  \\
3 & Continuum     & 3840  & 358.008 &1875.000  & 488.281  \\
\hline
\end{tabular}
\end{table}

\begin{table}
\small
\centering
\caption{Correlator Setup for Cycle 3 Project}
\label{tab:corr3}
\begin{tabular}{llllll}
\hline
Spectral  & Line or   & Number of & Central Frequency & Bandwidth & Channel Width\\
Window & Continuum & Channels  & (GHz)             & (MHz)     & (kHz) \\
\hline\hline
0 & SO \SOt      & 960   & 346.528 & 234.375  & 244.140  \\
1 & CO $J=3-2$      & 960   & 345.796 & 234.375  & 244.140  \\
2 & H$^{13}$CO$^+$ $J=4-3$      & 960   & 346.998 & 234.375  & 244.140  \\
3 & SiO $J=8-7$     & 960   & 347.330 & 234.375  & 244.140  \\
4 & HCO$^+$ $J=4-3$ & 1920  & 356.735 & 468.750  & 244.140  \\
5 & Continuum     & 1920  & 357.994 &1875.000  & 976.562  \\
\hline
\end{tabular}
\end{table}

\begin{table}
%\scriptsize
\small
\centering
\caption{Calibrators and Their Flux Densities}
\label{tab:calib}
\begin{tabular}{llll}
\hline
Date & Bandpass Calibrator &Flux  Calibrator & Phase Calibrator \\
(YYYY-MM-DD) & (Quasar, Flux Density) & (Quasar, Flux Density) & (Quasar, Flux Density) \\
\hline\hline
2015-08-29 & J0607-0834, 1.20 Jy & J0423-013, 1.03 Jy & J0552+0313, 0.25 Jy  \\ %X2e9e
2015-08-29 & J0607-0834, 1.20 Jy & J0423-013, 1.03 Jy &  J0552+0313, 0.25 Jy \\ %X26d4
2015-11-05 & J0423-0120, 0.55 Jy & J0423-0120, 0.55 Jy & J0541-0211, 0.22 Jy \\  %X877
2015-12-03 & J0510+1800, 4.07 Jy & J0423-0120, 0.67 Jy & J0541-0211, 0.23 Jy \\ %X3af
\hline
\end{tabular}
\end{table}

%A single pointing was used to observe the center in this system.

%The spectral resolution for the first 3 windows was set to have a velocity
%resolution of $\sim$ 0.2 \vkm{} per channel.
%Since \HCOP{} also traces the jet close to the central source, we will
%discuss the jet there too.  Observational results in CO and SiO will be
%presented in next paper to discuss the jet further out from the central
%source.

%, with Quasars J0538-440 and
%J0607-085 as passband calibrators, Quasar J0552+0313, J0541-0211 as a gain
%calibrator, and Callisto and Ganymede as flux calibrators.  

%The rms noise level is 0.61 \mJyb{} (i.e., 27 mK) for the continuum, and
%$\sim$ 10 \mJyb{} (i.e., 0.45 K) for the \HCOP{} channel maps.  The
%velocities in the channel maps are LSR.  

%The systemic velocity in this
%region is assumed to be $\Vsys= 1.7\pm0.1$ \vkm{} LSR.
%Throughout this paper, we define an offset velocity $\Voff = \VLSR - \Vsys$
%to facilitate our presentation.

\def\no{n_\mathrm{o}}
\def\na{n_\mathrm{t}}
\def\Ro{R_\mathrm{o}}
\def\Ra{R_\mathrm{t}}
\def\To{T_\mathrm{o}}
\def\Ta{T_\mathrm{t}}
\def\ho{h_\mathrm{o}}
\def\ha{h_\mathrm{t}}

\def\vk{v_\mathrm{ko}}
\def\cs{c_\mathrm{s}}
\def\vko{v_\mathrm{ko}}
\def\cso{c_\mathrm{so}}
\def\vp{v_\phi}

\subsection{Model} 

% 
%where $\Ro$ is the outer disk radius to be derived and $\To$ is the temperature
%at the outer disk radius.

In this section, we introduce a parametrized model for the disk emission, in
order to illustrate the formation of the dark lane sandwiched between two
brighter regions.

For a disk in vertical hydrostatic equilibrium, the scale height $h$
in the cylindrical coordinate system is given by\cite{Armitage2015}
\begin{equation}
\frac{h}{R} \sim \frac{\cs}{\vp} \propto \frac{R^{-q/2}}{R^{-1/2}} = 
(\frac{R}{R_o})^{(1-q)/2}
\end{equation}
where $R$ is the cylindrical radius and $c_s$ is the isothermal sound speed
proportional to $T^{1/2}$, where $T$ is the temperature.
Assuming $T \propto R^{-q}$, we have $c_s \propto R^{-q/2}$.
For reference, $q$ is $\sim$ 0.5 in
Taurus disks\cite{Andrews2009} and 0.75 in geometrically thin accretion disk
model\cite{Armitage2015}. $\vp$ is the rotational velocity assumed
to be Keplerian and thus proportional to $R^{-1/2}$, as suggested 
by C$^{17}$O gas kinematics\cite{Codella2014}.
In this case, the scale height of the disk increases with the disk radius monotonically.
However, the continuum map in Figure 1d appears to indicate that the disk
becomes thinner near the outer edge.  The physical reason for the behavior
is unclear; it could be due to self-shielding\cite{Dullemond2004} 
or optical depth effects.
To model the decrease of disk thickness near the edge, we introduce a
radius, $R_t$, beyond which the scale height
decreases exponentially. Let $\ha$ be the scale height of the disk at $\Ra$,
we have
\begin{eqnarray}
h (R) \sim \ha  \left\{ \begin{array}{ll}
(\frac{R}{\Ra})^{1+(1-q)/2}  & \;\;\textrm{if}\;\; R < \Ra,  \\ 
\exp[-(\frac{R-\Ra}{\Ro-\Ra})^2] & \;\;\textrm{if}\;\; \Ra \leq R \leq \Ro
\end{array}  \right.
\label{eq:thick}
\end{eqnarray}
where $\Ro$ is the outer radius of the disk.

%The disk prescribed by Equation \ref{eq:thick} is illustrated in Figure 3a.

% (e.g., Qi et al.  2011, ApJ, 740, 84).

The temperature increases not only radially toward the center but also vertically 
from the disk midplane to the surface for a passive disk (assumed here),
either because of radiative heating by the central source
or mechanical heating from wind-disk interaction. 
The detailed temperature structure depends on many factors, such as the
properties of the dust grains and their spatial distribution, which are
uncertain.
For illustrative purposes, we adopt
\begin{eqnarray}
T  \sim \Ta (\frac{R}{\Ra})^{-q} \left\{ \begin{array}{ll}
\exp(\frac{z^2}{2h^2})  & \;\;\textrm{if}\;\; R < \Ra,  \\ 
1 & \;\;\textrm{if}\;\; \Ra \leq R \leq \Ro
\end{array}  \right.
\end{eqnarray}
where $\Ta$ is the temperature in the disk midplane at $\Ra$.
The number density of molecular hydrogen is assumed to be\cite{Armitage2015}
\begin{equation}
n= \na (\frac{R}{\Ra})^{-p} \exp(-\frac{z^2}{2 h^2})
\end{equation}
with the power-law index $p=2$ and $\na$ being the number density in the disk midplane
at $\Ra$.
Helium is included so that the mass density is $\rho=1.4 n \mH2$.

%Note that $h$ defined here is $\sqrt{2}$ times the disk scale height $R c_s/v_\phi$.

{We used our radiative transfer code to obtain dust continuum emission map
from the model disk.  We first computed the thermal emission from each point in
the disk based on its local temperature, and then generated a synthetic map
by integrating along each line of sight the local emission that is
attenuated by the optical depth. With the dust opacity spectral index
$\beta\sim0.6$ found in previous observations of the HH 212
disk\cite{Lee2008} and the dust opacity formula for circumstellar
disks\cite{Beckwith1990}, we derived a dust opacity $\kappa \sim 0.054$
cm$^2$ g$^{-1}$ at 850 \micron{}.} We assume $p=2$ and $q=0.75$ for the
disk\cite{Armitage2015}.  Based on the disk structure seen in the dust
emission map in Figure 1d, we find that $\Ro \sim$
\arcsa{0}{17}$\pm$\arcsa{0}{03} (68$\pm$12 AU), $\Ra \sim$
\arcsa{0}{09}$\pm$\arcsa{0}{02} (36$\pm$8 AU), and $\ha \sim$
\arcsa{0}{03}$\pm$\arcsa{0}{01} (12$\pm$4 AU).  {To produce the image
shown in Figure 3, we adopted $\na \sim 2.2\pm0.6\times10^{10}$ \cmc{}} and
$\Ta\sim$ 73$\pm$10 K.  
With this value of $n_t$, the emission becomes optically thick at $R\sim $
\arcsa{0}{1} (40 AU).  The value of $T_t$ yields a flux density of $\sim
2.0$~\mJyb{} in a \arcsa{0}{02} beam (or a brightness temperature of
$\sim$ 55 K) in the optically thick region,
consistent with Figure 2. The disk has a total mass of 
\begin{equation} 
M_D= 1.4\,\mH2 \int n\,2\pi\,R\,dR\,dz \sim 0.05\,\solarmasse 
\end{equation} 
Since part of the observed emission comes from the dust
thermal emission and part from the dust scattering of the thermal emission, 
the disk mass derived here is likely an upper limit.
Note that the mid-plane temperature of 73~K at the radius $R_t=36$~AU yields
an isothermal scale height of $\sim 8.3$~AU for an estimated central mass of
0.2 M$_\odot$.  This value is somewhat smaller than the best fit value of
$h_t\sim 12$~AU.  The agreement between the two may be improved when the
variation of temperature with height is taken into account.  More detailed
treatment of heating and cooling is needed to draw a firmer conclusion.


\begin{thebibliography}{}

\bibitem[1]{Simon2000} M.~Simon, A.~Dutrey, S.~Guilloteau, Dynamical
Masses of T Tauri Stars and Calibration of Pre-Main-Sequence Evolution.\
\apj, {\bf 545}, 1034-1043 (2000).


\bibitem[2]{Perez2016} L.~M.~P{\'e}rez, J.~M.~Carpenter, S.~M.~Andrews,
L.~Ricci, A.~Isella, H.~Linz, A.~I.~Sargent, D.~J.~Wilner, T.~Henning,
A.~T.~Deller, C.~J.~Chandler, C.~P.~Dullemond, J.~Lazio, K.~M.~Menten,
S.~A.~Corder, S.~Storm, L.~Testi, M.~Tazzari, W.~Kwon, N.~Calvet,
J.~S.~Greaves, R.~J.~Harris, L.~G.~Mundy, Spiral density waves in a young
protoplanetary disk.\ {\it Science}, {\bf 353}, 1519-1521 (2016).


\bibitem[3]{Lee2009} C.-F.~Lee, N.~Hirano, A.~Palau, P.~T.~P.~Ho,
T.~L.~Bourke, Q.~Zhang, H.~Shang, Rotation and Outflow Motions in the
Very Low-Mass Class 0 Protostellar System HH 211 at Subarcsecond
Resolution. \apj, {\bf 699}, 1584-1594 (2009).


\bibitem[4]{Tobin2012} J.~J.~Tobin, L.~Hartmann, H.-F.~Chiang, D.~J.~Wilner,
L.~W.~Looney, L.~Loinard, N.~Calvet, P.~D'Alessio, A \~{}0.2-solar-mass
protostar with a Keplerian disk in the very young L1527 IRS system. \nat,
{\bf 492}, 83-85 (2012).

\bibitem[5]{Murillo2013} N.~M.~Murillo, S.-P.~Lai, S.~Bruderer, D.~Harsono,
E.~F.~van Dishoeck, A Keplerian disk around a Class 0 source: ALMA
observations of VLA1623A.\ \aap, {\bf 560}, A103(2013).

\bibitem[6]{Terebey1984} S.~Terebey, F.~H.~Shu, P.~Cassen,  The
collapse of the cores of slowly rotating isothermal clouds.\ \apj, {\bf
286}, 529-551 (1984).

\bibitem[7]{Chapman2013} N.~L.~Chapman, J.~A.~Davidson, P.~F.~Goldsmith,
M.~Houde, W.~Kwon, Z.-Y.~Li, L.~W.~Looney, B.~Matthews, T.~G.~Matthews,
G.~Novak, R.~Peng, J.~E.~Vaillancourt, N.~H.~Volgenau, Alignment between
Flattened Protostellar Infall Envelopes and Ambient Magnetic Fields.\ \apj,
{\bf 770}, 151-164 (2013).

\bibitem[8]{Li2014} Z.-Y.~Li, R.~Banerjee, R.~E.~Pudritz,
J.~K.~J{\o}rgensen, H.~Shang, R.~Krasnopolsky, A.~Maury, The Earliest
Stages of Star and Planet Formation: Core Collapse, and the Formation of
Disks and Outflows.\ Protostars and Planets, {\bf VI}, 173-194 (2014).

\bibitem[9]{Allen2003} A.~Allen, Z.-Y.~Li, F.~H.~Shu,  Collapse of
Magnetized Singular Isothermal Toroids.  II.  Rotation and Magnetic
Braking.\ \apj, {\bf 599}, 363-379 (2003).

\bibitem[10]{Joos2012} M.~Joos, P.~Hennebelle, A.~Ciardi,  Protostellar
disk formation and transport of angular momentum during magnetized core
collapse.\ \aap, {\bf 543}, A128-A149 (2012).

\bibitem[11]{Li2013} Z.-Y.~Li, R.~Krasnopolsky, H.~Shang, Does
Magnetic-field-Rotation Misalignment Solve the Magnetic Braking Catastrophe
in Protostellar Disk Formation?\ \apj, {\bf 774}, 82-93 (2013).

\bibitem[12]{Tomida2015} K.~Tomida, S.~Okuzumi, M.~N.~Machida,
Radiation Magnetohydrodynamic Simulations of Protostellar Collapse: Nonideal
Magnetohydrodynamic Effects and Early Formation of Circumstellar Disks.\
\apj, {\bf 801}, 117-136 (2015).

\bibitem[13]{Shu2008} F.~H.~Shu, S.~Lizano, D.~Galli, M.~J.~Cai,
S.~Mohanty, The Challenge of Sub-Keplerian Rotation for Disk Winds.\ \apjl,
{\bf 682}, L121-L124 (2008).


\bibitem[14]{Lee2014} C.-F.~Lee, N.~Hirano, Q.~Zhang, H.~Shang, P.~T.~P.~Ho,
R.~Krasnopolsky, ALMA Results of the Pseudodisk, Rotating Disk, and Jet
in the Continuum and HCO$^{+}$ in the Protostellar System HH 212.\ \apj,
{\bf 786}, 114-125 (2014).

\bibitem[15]{Codella2014} C.~Codella, S.~Cabrit, F.~Gueth, L.~Podio,
S.~Leurini, R.~Bachiller, A.~Gusdorf, B.~Lefloch, B.~Nisini, M.~Tafalla,
W.~Yvart, The ALMA view of the protostellar system HH212.  The wind, the
cavity, and the disk.\ \aap, {\bf 568}, L5-L9 (2014).


\bibitem[16]{Zinnecker1992} H.~Zinnecker, P.~Bastien, J.-P.~Arcoragi,
H.~W.~Yorke, Submillimeter dust continuum observations of three low
luminosity protostellar IRAS sources.\ \aap, {\bf 265}, 726-732 (1992).

\bibitem[17]{Zinnecker1998} H.~Zinnecker, M.~J.~McCaughrean, and
J.~T.~Rayner, A symmetrically pulsed jet of gas from an invisible protostar
in Orion.\ \nat, {\bf 394}, 862-865 (1998).


\bibitem[18]{Konigl2000} A.~Konigl, R.~E.~Pudritz,  Disk Winds and the
Accretion-Outflow Connection.\ Protostars and Planets, {\bf IV}, 759(2000).

\bibitem[19]{Shu2000} F.~H.~Shu, J.~R.~Najita, H.~Shang, Z.-Y.~Li,
X-Winds Theory and Observations.\ Protostars and Planets, {\bf IV}, 789-814
(2000).

\bibitem[20]{McCaughrean2002} M.~McCaughrean, H.~Zinnecker, M.~Andersen,
G.~Meeus, N.~Lodieu, Standing on the shoulder of a giant: ISAAC, Antu,
and star formation.\ The Messenger, {\bf 109}, 28-36 (2002).

\bibitem[21]{Lee2015} C.-F.~Lee, N.~Hirano, Q.~Zhang, H.~Shang,
P.~T.~P.~Ho, Y.~Mizuno, Jet Motion, Internal Working Surfaces, and
Nested Shells in the Protostellar System HH 212.\ \apj, {\bf 805}, 186-194
(2015).

\bibitem[22]{Wiseman2001} J.~Wiseman, A.~Wootten, H.~Zinnecker, and
M.~McCaughrean, The Flattened, Rotating Molecular Gas Core of Protostellar
Jet HH 212.\ \apjl, {\bf 550}, L87-L90 (2001).

\bibitem[23]{Krasnopolsky2002} R.~Krasnopolsky, A.~K{\"o}nigl,
Self-similar Collapse of Rotating Magnetic Molecular Cloud Cores.\ \apj,
{\bf 580}, 987-1012 (2002).

\bibitem[24]{Claussen1998} M.~J.~Claussen, K.~B.~Marvel, A.~Wootten, and
B.~A.~Wilking, Distribution and Motion of the Water Masers near IRAS
05413-0104.\ \apjl, {\bf 507}, L79-L82 (1998).

\bibitem[25]{Lee2007} C.-F.~Lee, P.~T.~P.~Ho, N.~Hirano, H.~Beuther,
T.~L.~Bourke, H.~Shang, Q.~Zhang, HH 212: Submillimeter Array
Observations of a Remarkable Protostellar Jet.\ \apj, {\bf 659}, 499-511
(2007).

\bibitem[26]{Chiang1997} E.~I~Chiang, P.~Goldreich,  Spectral Energy
Distributions of T Tauri Stars with Passive Circumstellar Disks.\ \apj, {\bf
490}, 368-376 (1997).

\bibitem[27]{Li1996} Z.-Y.~Li, F.~H.~Shu,  Interaction of Wide-Angle MHD
Winds with Flared Disks.\ \apj, {\bf 468}, 261-268 (1996).

\bibitem[28]{Lee2008} C.-F.~Lee, P.~T.~P.~Ho, T.~L.~Bourke, N.~Hirano,
H.~Shang, Q.~Zhang, SiO Shocks of the Protostellar Jet HH 212: A Search
for Jet Rotation.\ \apj, {\bf 685}, 1026-1032 (2008).

\bibitem[29]{Natta2007} Natta, A., L.~Testi, N.~Calvet, T.~Henning,
R.~Waters, and D.~Wilner, Dust in Protoplanetary Disks: Properties and
Evolution.\ Protostars and Planets {\bf V}, 767-781 (2007).


\bibitem[30]{Draine2006} B.~T.~Draine,  On the Submillimeter Opacity of
Protoplanetary Disks.\ \apj, {\bf 636}, 1114-1120 (2006).

\bibitem[31]{Kataoka2015} A.~Kataoka, T.~Muto, M.~Momose, T.~Tsukagoshi,
M.~Fukagawa, H.~Shibai, T.~Hanawa, K.~Murakawa, C.~P.~Dullemond,
Millimeter-wave Polarization of Protoplanetary Disks due to Dust
Scattering.\ ,\apj, {\bf 809}, 78-92 (2015).

\bibitem[32]{Beckwith1990} S.~V.~W.~Beckwith, A.~I.~Sargent, R.~S.~Chini,
R.~Guesten, A survey for circumstellar disks around young stellar
objects.\ \aj, {\bf 99}, 924-945 (1990).

\bibitem[32a]{Li2017} J.I-H~Li,H.B.~Liu,Y.~Hasegawa,N.~Hirano, Systematic
analysis of SED and the dust opacity indices for Class 0 YSOs.\ \apj, {\it
accepted}


\bibitem[33]{Nakano2002} T.~Nakano, R.~Nishi, T.~Umebayashi,  Mechanism
of Magnetic Flux Loss in Molecular Clouds.\ \apj, {\bf 573}, 199-214 (2002).


\bibitem[34]{Armitage2015} P.~J.~Armitage,  Physical processes in
protoplanetary disks.\ , arXiv:1509.06382(2015).

\bibitem[35]{Andrews2009} S.~M.~Andrews, D.~J.~Wilner, A.~M.~Hughes, C.~Qi,
C.~P.~Dullemond, Protoplanetary Disk Structures in Ophiuchus.\ \apj,
{\bf 700}, 1502-1523 (2009).

\bibitem[36]{Dullemond2004} C.~P.~Dullemond, C.~Dominik,  Flaring vs.
self-shadowed disks: The SEDs of Herbig Ae/Be stars.\ \aap, {\bf 417},
159-168 (2004).



%\bibitem[2]{Choi2010} Choi, M., Tatematsu, K., \& Kang, M.\ 2010, \apjl, 723, L34 


%\bibitem[Alten et al.(1997)]{Alten1997} 
% Alten, V. P., Bally, J., Devine, D. and Miller, G. J. 1997, IAU Symp. 182:
% Herbig-Haro Flows and the Birth of Stars, 182, 51P
%\bibitem[Anderson et al.(2003)]{Anderson2003} Anderson, J.~M., Li, 
%Z.-Y., Krasnopolsky, R., \& Blandford, R.~D.\ 2003, \apjl, 590, L107 
%\bibitem[Andr\'e \& Montmerle(1994)]{Andre1994} Andr\'e, P.  \& 
%  Montmerle, T. 1994, \apj, 420, 837 
%\bibitem[Andre et al.(2000)]{Andre2000} Andre, P., Ward-Thompson, 
%D., \& Barsony, M.\ 2000, Protostars and Planets IV, 59  
%\bibitem[Andrews 
%\& Williams(2007)]{Andrews2007} Andrews, S.~M., \& Williams, J.~P.\
%2007, \apj, 659, 705 
%\bibitem[Anglada \& Rodr{\'{\i}}guez(2002)]{Anglada2002} Anglada, 
%G.~\& Rodr{\'{\i}}guez, L.~F.\ 2002, Revista Mexicana de Astronomia y 
% Astrofisica, 38, 13 
%\bibitem[Arce \& Sargent(2004)]{Arce2004} Arce, H.~G.~\& 
% Sargent, A.~I.\ 2004, \apj, 612, 342 
%\bibitem[Arce et al.(2007)]{Arce2007} Arce, H.~G., Shepherd, D., 
%Gueth, F., Lee, C.-F., Bachiller, R., Rosen, A., 
%\& Beuther, H.\ 2007, Protostars and Planets V, 245 
%\bibitem[Bachiller(1996)]{Bachiller1996}Bachiller, R. 1996, ARAA, 34, 111
%\bibitem[Bachiller et al.(1995)]{Bachiller1995} Bachiller, R., 
% Guilloteau, S., Dutrey, A., Planesas, P. \& Martin-Pintado, J. 1995, \aap, 
% 299, 857 
%\bibitem[Bally \& Lada(1983)]{Bally1983} Bally, J. \& Lada, C. J. 
% 1983, \apj, 265, 824 
%\bibitem[Belloche \& Andr{\' e}(2004)]{Belloche2004} Belloche, A.~\& 
% Andr{\' e}, P.\ 2004, \aap, 419, L35 
%\bibitem[Bence, Richer \& Padman(1996)]{Bence1996} Bence, S. J., 
%  Richer, J. S. \& Padman, R. 1996, \mnras, 279, 866 
%\bibitem[Bence et al.(1998)]{Bence1998} Bence, S. J., Padman, R., 
% Isaak, K. G., Wiedner, M. C. \& Wright, G. S. 1998, \mnras, 299, 965 
%\bibitem[Bohigas, Persi \& Tapia(1993)]{Bohigas1993} Bohigas, J., 
% Persi, P. \& Tapia, M. 1993, \aap, 267, 168 
%\bibitem[Cabrit, Raga \& Gueth(1997)]{Cabrit1997} Cabrit, S., Raga, A., Gueth, F.
%  1997, IAUS, 182, 163
%\bibitem[Cabrit et al.(2007)]{Cabrit2007} Cabrit, S., Codella, C., 
%Gueth, F., Nisini, B., Gusdorf, A., Dougados, C., \& Bacciotti, F.\ 2007, 
%\aap, 468, L29 
%\bibitem[Cabrit et 
%al.(2012)]{Cabrit2012} Cabrit, S., Codella, C., Gueth, F., \& Gusdorf, A.\ 2012, \aap, 548, L2 
%\bibitem[Caselli, Myers, \& Thaddeus(1995)]{Caselli1995} Caselli, 
% P., Myers, P.~C., \& Thaddeus, P.\ 1995, \apjl, 455, L77 
%\bibitem[Caselli et al.(1997)]{Caselli1997} Caselli, P., Hartquist, 
%T.~W., \& Havnes, O.\ 1997, \aap, 322, 296 
%\bibitem[Caselli et al.(2002)]{Caselli2002} Caselli, P., Benson, 
%P.~J., Myers, P.~C., \& Tafalla, M.\ 2002, \apj, 572, 238
%\bibitem[Cerqueira et 
%al.(2006)]{Cerqueira2006} Cerqueira, A.~H., Vel{\'a}zquez, P.~F.,
%Raga, A.~C., Vasconcelos, M.~J., \& de Colle, F.\ 2006, \aap, 448, 231 
%\bibitem[Chapman et al.(2002)]{Chapman2002} Chapman, N.~L., Mundy, 
%L.~G., Lee, C.-F., \& White, S.~M.\ 2002, Bulletin of the American 
%Astronomical Society, 34, 1133 
%\bibitem[Chernin et al.(1994)]{Chernin1994} Chernin, L.M., Masson, C.R.,
%  Gouveia dal Pino, E.M., Benz, W., 1994, ApJ, 426, 204
%\bibitem[Chini et al.(1997)]{Chini1997} Chini, R., Reipurth, B., 
%Sievers, A., Ward-Thompson, D., Haslam, C.~G.~T., Kreysa, E., \& Lemke, R.\ 
%1997, \aap, 325, 542 


%\bibitem[Cernicharo \& Reipurth(1996)]{Cernicharo1996} Cernicharo, J. 
%  \& Reipurth, B. 1996, \apjl, 460, L57 
%\bibitem[Chrysostomou et al.(2008)]{Chrysostomou2008} Chrysostomou, A., 
%Bacciotti, F., Nisni, B., Ray, T.~P., Eisloffel, J., Davis, C.~J., 
%\& Takami, M. \ 2008, \aap, 482, 575 
%\bibitem[Codella et al.(2005)]{Codella2005} Codella, C., Bachiller, 
%R., Benedettini, M., Caselli, P., Viti, S., \& Wakelam, V.\ 2005, \mnras, 
%361, 244 
%\bibitem[Codella et al.(2007)]{Codella2007} Codella, C., Cabrit, 
%S., Gueth, F., Cesaroni, R., Bacciotti, F., Lefloch, B., \& McCaughrean, 
%M.~J.\ 2007, \aap, 462, L53 

%\bibitem[Codella et al.(2016)]{Codella2016} Codella, C., Ceccarelli, C.,
%Cabrit, S., et al.\ 2016, \aap, 586, L3

%\bibitem[Coffey et al.(2007)]{Coffey2007} Coffey, D., Bacciotti, 
%F., Ray, T.~P., Eisl{\"o}ffel, J., \& Woitas, J.\ 2007, \apj, 663, 350 
%\bibitem[Cohen(1980)]{Cohen1980} Cohen, M. 1980, \aj, 85, 29 
%\bibitem[Coppin, Davis \& Micono(1998)]{Coppin1998} Coppin, K. E. 
%  K., Davis, C. J. \& Micono, M.  1998, \mnras, 301, L10 
%\bibitem[Davis, Mundt \& Eisloeffel(1994)]{Davis1994} Davis, C. J., 
%  Mundt, R.  \& Eisloeffel, J.  1994, \apjl, 437, L55 
%\bibitem[Davis et al.(1997)]{Davis1997} Davis, 
%  C. J., Ray, T. P., Eisloeffel, J. \& Corcoran, D. 1997, \aap, 324, 263 
%\bibitem[Davis et al.(2000)]{Davis2000} Davis, C.~J., Berndsen, 
%A., Smith, M.~D., Chrysostomou, A., \& Hobson, J.\ 2000, \mnras, 314, 241 
%\bibitem[de Geus, Bronfman, \& Thaddeus(1990)]{deGeus1990} de 
% Geus, E.~J., Bronfman, L., \& Thaddeus, P.\ 1990, A\&A, 231, 137 
%\bibitem[Delamarter, Frank, \& Hartmann(2000)]{Delamarter2000} 
% Delamarter, G., Frank, A., \& Hartmann, L.\ 2000, \apj, 530, 923
%\bibitem[Dent et al.(2003)]{Dent2003} Dent, W.~R.~F., Furuya, 
%R.~S., \& Davis, C.~J.\ 2003, \mnras, 339, 633 
%\bibitem[Downes \& Ray(1999)]{Downes1999} Downes, T. P.  Ray, T. P. 
%1999, \aap, 345, 977 


%\bibitem[Evans(1999)]{Evans1999} Evans, N.~J., II 1999, \araa, 37, 311 
%\bibitem[Fendt \& Zinnecker(1998)]{Fendt1998} Fendt, C., \& 
%Zinnecker, H.\ 1998, \aap, 334, 750 
%\bibitem[Frerking et al.(1982)]{Frerking1982} Frerking, M.~A., 
% Langer, W.~D., \& Wilson, R.~W.\ 1982, \apj, 262, 590 
%\bibitem[Fukui et al.(1993)]{Fukui1993} Fukui, Y. , Iwata, T. , 
% Mizuno, A. , Bally, J.  \& Lane, A. P. 1993, Protostars and planets III 
% (A93-42937 17-90), p. 603-639., 603 
%\bibitem[Galli \& Shu(1993)]{Galli1993} Galli, D., \& Shu, F.~H.\ 
% 1993, \apj, 417, 243 
%\bibitem[Galv{\' a}n-Madrid et al.(2004)]{Galvan2004} Galv{\' 
%a}n-Madrid, R., Avila, R., \& Rodr{\'{\i}}guez, L.~F.\ 2004, Revista 
%Mexicana de Astronomia y Astrofisica, 40, 31 
%\bibitem[Garnavich et al.(1997)]{Garnavich1997} 
%  Garnavich, P. M., Noriega-Crespo, A. , Raga, A. C. \& Bohm, K. -H.  1997, 
%  \apj, 490, 752
%\bibitem[Gibb et al.(2004)]{Gibb2004} Gibb, A.~G., Richer, 
%J.~S., Chandler, C.~J., \& Davis, C.~J.\ 2004, \apj, 603, 198 
%\bibitem[Girart et al.(2000)]{Girart2000} Girart, J.~M., 
% Estalella, R., Ho, P.~T.~P., \& Rudolph, A.~L.\ 2000, \apj, 539, 763 
%\bibitem[Glassgold et al.(1991)]{Glassgold1991} Glassgold, A.~E., 
%Mamon, G.~A., \& Huggins, P.~J.\ 1991, \apj, 373, 254 
%\bibitem[de Gouveia dal Pino \& Benz (1993)]{Gouveia1993} de Gouveia 
% dal Pino, E. M. \& Benz, W.  1993, \apj, 410, 686  
%\bibitem[Gredel \& Reipurth(1994)]{Gredel1994} Gredel, R.  \&
% Reipurth, B. 1994, \aap, 289, L19 
%\bibitem[Gottlieb, Myers, \& Thaddeus(2003)]{Gottlieb2003} Gottlieb, 
% C.~A., Myers, P.~C., \& Thaddeus, P.\ 2003, \apj, 588, 655 
%\bibitem[Gueth et al.(1997)]{Gueth1997} Gueth, F., Guilloteau, S., Dutrey, 
% A., \& Bachiller, R.\ 1997, \aap, 323, 943 
%\bibitem[Gueth et al.(1998)]{Gueth1998} Gueth, F., Guilloteau, 
%S., \& Bachiller, R.\ 1998, \aap, 333, 287 
%\bibitem[Gueth \& Guilloteau(1999)]{Gueth1999} Gueth, F. \& 
%  Guilloteau, S. 1999, \aap, 343, 571 
%\bibitem[Gueth, Guilloteau \& Bachiller(1996)]{Gueth1996} Gueth, 
%  F., Guilloteau, S. \& Bachiller, R. 1996, \aap, 307, 891
%\bibitem[Harsono et al.(2013)]{Harsono2013} Harsono, D., Jorgensen, 
%J.~K., van Dishoeck, E.~F., et al.\ 2013, arXiv:1312.5716 
%\bibitem[Hartigan et al.(2000)]{Hartigan2000} Hartigan, P., Bally,
%J., Reipurth, B., \& Morse, J.~A.\ 2000, Protostars and Planets IV, 841
%\bibitem[Hartmann et al.(1994)]{Hartmann1994} Hartmann, L., Boss, 
% A., Calvet, N., \& Whitney, B.\ 1994, \apjl, 430, L49 
%\bibitem[Hartmann et al.(1996)]{Hartmann1996} Hartmann, L., Calvet, 
%N., \& Boss, A.\ 1996, \apj, 464, 387
%\bibitem[Hayashi et al.(1993)]{Hayashi1993} Hayashi, M., Ohashi, 
%N., \& Miyama, S.~M.\ 1993, \apjl, 418, L71 
%\bibitem[Hirano et al.(2006)]{Hirano2006} Hirano, N., Liu, S.-Y., 
%Shang, H., Ho, P.~T.~P., Huang, H.-C., Kuan, Y.-J., McCaughrean, M.~J., \& 
%Zhang, Q.\ 2006, \apjl, 636, L141 
%\bibitem[Ho et al.(2004)]{Ho2004} Ho, P. T. P., Moran, J. M., \& Lo, K. Y. 
%B2004, ApJ, 616, L1
%\bibitem[Hodapp(1994)]{Hodapp1994} Hodapp, K. -W.  1994, \apjs, 94, 
% 615 
%\bibitem[Hodapp \& Ladd(1995)]{Hodapp1995} Hodapp, K. -W.  \& Ladd, 
%  E. F. 1995, \apj, 453, 715  
%\bibitem[Hogerheijde et al.(1998)]{Hogerheijde1998} Hogerheijde, M.~R., 
%van Dishoeck, E.~F., Blake, G.~A., \& van Langevelde, H.~J.\ 1998, \apj, 
%B502, 315
%\bibitem[Hogerheijde(2001)]{Hogerheijde2001} Hogerheijde, M.~R.\ 2001, 
%\apj, 553, 618
%\bibitem[J{\o}rgensen et al.(2004)]{Jorgensen2004} J{\o}rgensen, 
% J.~K., Sch{\" o}ier, F.~L., \& van Dishoeck, E.~F.\ 2004, \aap, 416, 603 
%\bibitem[J{\o}rgensen et al.(2007)]{Jorgensen2007} J{\o}rgensen, 
%J.~K., et al.\ 2007, \apj, 659, 479 
%\bibitem[Keto et al.(1988)]{Keto1988} Keto, E.~R., Ho, P.~T.~P., 
%\& Haschick, A.~D.\ 1988, \apj, 324, 920 
%\bibitem[Keto 
%\& Zhang(2010)]{Keto2010} Keto, E., \& Zhang, Q.\ 2010, \mnras, 406, 102 
%\bibitem[Krasnopolsky \& K\"{o}nigl(2002)]{Krasnopolsky2002} Krasnopolsky,
%R., \& K\"{o}nigl, A.\ 2002, \apj, 580, 987

%\bibitem[Kumar, Anandarao \& Davis(1999)]{Kumar1999} Kumar, M. S. 
% N. , Anandarao, B. G. \& Davis, C. J. 1999, \aap, 344, L9 
%\bibitem[Lada(1985)]{Lada1985} Lada, C. J. 1985, \araa, 23, 267 
%\bibitem[Lada \& Fich(1996)]{Lada1996} Lada, C. J. \& Fich, M.  
%  1996, \apj, 459, 638 
%\bibitem[Lee(2010)]{Lee2010} Lee, C.-F.\ 2010, \apj, 725, 712 
%\bibitem[Lee(2011)]{Lee2011} Lee, C.-F.\ 2011, \apj, 741, 62 

%\bibitem[Lee et al.(2000)]{Lee2000} Lee, C.-F., Mundy, L.G., Reipurth, B.,
%  Ostriker, E.C., \& Stone, J.M. 2000, \apj, 542, 925
%\bibitem[Lee et al.(2001)]{Lee2001} Lee, 
% C.-F., Stone, J.~M., Ostriker, E.~C., \& Mundy, L.~G.\ 2001, \apj, 557, 429  
%\bibitem[Lee et al.(2006)]{Lee2006} Lee, C.-F., Ho, P.~T.~P., 
%Beuther, H., Bourke, T.~L., Zhang, Q., Hirano, N., \& Shang, H.\ 2006,
%\apj, 639, 292
%\bibitem[Lee et al.(2007b)]{Lee2007b} Lee, C.-F., Ho, P.~T.~P., 
%Palau, A., Hirano, N., Bourke, T.~L., Shang, H., 
%\& Zhang, Q.\ 2007b, \apj, 670, 1188 
%\bibitem[Lee et al.(2008)]{Lee2008} Lee, C.-F., Ho, P.~T.~P., 
%Bourke, T.~L., et al.\ 2008, \apj, 685, 1026 
%\bibitem[Lee, Myers, \& Tafalla(2001)]{LMT2001} Lee, C.~W., 
% Myers, P.~C., \& Tafalla, M.\ 2001, \apjs, 136, 703
%\bibitem[Lee et al.(2002)]{Lee2002} Lee, 
% C.-F., Mundy, L.~G., Stone, J.~M., \& Ostriker, E.~C.\ 2002, \apj, 576, 294 
%\bibitem[Lee et al.(2005)]{Lee2005a} Lee, C., Ho, P.~T.~P., \& 
% White, S.~M.\ 2005, \apj, 619, 948
%\bibitem[Lee \& Ho(2005)]{Lee2005b} Lee, C., \& Ho, P.~T.~P.,
%\ 2005, \apj, in press
%\bibitem[Lee \& Sahai(2003)]{Lee2003} Lee, C.-F.~\& Sahai, R.\ 
%2003, \apj, 586, 319 
%\bibitem[Levreault(1988)]{Levreault1988} Levreault, R. M. 1988, \apj, 
%  330, 897


%\bibitem[Lin et al.(1994)]{Lin1994} Lin, D.~N.~C., Hayashi, M., 
%Bell, K.~R., \& Ohashi, N.\ 1994, \apj, 435, 821
%\bibitem[Linke \& Goldsmith(1980)]{Linke1980} Linke, R.~A., \& 
% Goldsmith, P.~F.\ 1980, \apj, 235, 437  
%\bibitem[Li \& Shu(1996)]{Li1996} Li, Z. -Y.  \& Shu, F. H. 
%  1996, \apj, 472, 211 
%\bibitem[Mardones et al.(1997)]{Mardones1997} Mardones, D., Myers, 
%P.~C., Tafalla, M., Wilner, D.~J., Bachiller, R., \& Garay, G.\ 1997, \apj, 
%489, 719
%\bibitem[Marvel et al.(in press)]{Marvelpress}Marvel, K.B., McCaughrean, M.J., 
%  \& Sargent, A.I., manuscript in preparation
%\bibitem[Masson \& Chernin(1993)]{Masson1993} Masson, C. R. \& 
%  Chernin, L. M. 1993, \apj, 414, 230 
%\bibitem[Mathieu et al.(1988)]{Mathieu1988} Mathieu, R. D., Myers, 
%  P. C., Schild, R. E., Benson, P. J. \& Fuller, G. A. 1988, \apj, 330, 385 
%\bibitem[Matzner \& McKee(1999)]{Matzner1999} Matzner, C. D. \& 
%  McKee, C. F. 1999, \apjl, 526, L109 
%\bibitem[McKee et al.(1987)]{Mckee1987} McKee, 
%  C. F., Hollenbach, D. J., Seab, G. C. \& Tielens, A. G. G. M. 1987, \apj, 
%  318, 674
%\bibitem[Mellon \& Li(2008)]{Mellon2008} Mellon, R.~R., \& Li, Z.-Y.\ 2008, \apj,
%681, 1356 
%\bibitem[Menten et al.(2007)]{Menten2007} Menten, K.~M., Reid, M.~J., Forbrich, J., \&
%Brunthaler, A.\ 2007, \aap, 474, 515 
%\bibitem[Meyers-Rice \& Lada(1991)]{Meyers1991} Meyers-Rice, B. A. 
%\& Lada, C. J. 1991, \apj, 368, 445 
%\bibitem[Micono et al.(1998)]{Micono1998} Micono, M., Davis, C. J., 
%Ray, T. P., Eisloeffel, J. \& Shetrone, M. D. 1998, \apjl, 494, L227 
%\bibitem[Momose et al.(1998)]{Momose1998} Momose, M., Ohashi, N., 
% Kawabe, R., Nakano, T., \& Hayashi, M.\ 1998, \apj, 504, 314 
%\bibitem[Myers et al.(1987)]{Myers1987} Myers, P.~C., Fuller, 
%G.~A., Mathieu, R.~D., Beichman, C.~A., Benson, P.~J., Schild, R.~E., \& 
%Emerson, J.~P.\ 1987, \apj, 319, 340 
%\bibitem[Myers et al.(1988)]{Myers1988} Myers, 
%\bibitem[Myers \& Ladd(1993)]{Myers1993} Myers, P.~C.~\& Ladd, 
%E.~F.\ 1993, \apjl, 413, L47 
%P. C., Heyer, M., Snell, R. L. \& Goldsmith, P. F. 1988, \apj, 324, 907 
%\bibitem[Myers et al.(2000)]{Myers2000} Myers, P.~C., Evans, 
%N.~J., \& Ohashi, N.\ 2000, Protostars and Planets IV, 217 
%\bibitem[Nagar et al.(1997)]{Nagar1997} Nagar, N. 
% M., Vogel, S. N., Stone, J. M. \& Ostriker, E. C. 1997, \apjl, 482, L195 
%\bibitem[Ostriker(1997)]{Ostriker1997} Ostriker, E. C. 1997, \apj, 486, 291
%\bibitem[Nakamura(2000)]{Nakamura2000} Nakamura, F.\ 2000, \apj, 
%543, 291

%\bibitem[Ohashi et al.(1997)]{Ohashi1997} Ohashi, N., Hayashi, M., 
%Ho, P.~T.~P., \& Momose, M.\ 1997, \apj, 475, 211 
%\bibitem[Ostriker et. al.(2001)]{Ostriker2001} 
% Ostriker, E.~C., Lee, C., Stone, J.~M., \& Mundy, L.~G.\ 2001, \apj, 557, 
% 443 
%\bibitem[Palau et al.(2006)]{Palau2006} Palau, A., et al.\ 2006, 
%\apjl, 636, L137
%\bibitem[Pudritz et al.(2007)]{Pudritz2007} Pudritz, R.~E., Ouyed, 
%R., Fendt, C., \& Brandenburg, A.\ 2007, Protostars and Planets V, 277 
%\bibitem[Pirogov et al.(2003)]{Pirogov2003} Pirogov, L., Zinchenko, 
%I., Caselli, P., Johansson, L.~E.~B., \& Myers, P.~C.\ 2003, \aap, 405, 639 
%\bibitem[Raga et al.(1990)]{Raga1990} Raga, A.~C., Binette, L., 
%Canto, J., \& Calvet, N.\ 1990, \apj, 364, 601
%\bibitem[Raga (1993)]{Raga1993A} Raga, A. C. 1993, \apss, 208, 163 
%\bibitem[Raga \& Cabrit(1993)]{Raga1993} Raga, A. \& Cabrit, S. 
%1993, \aap, 278, 267 
%\bibitem[Raga et al.(1993b)]{Raga1993b} Raga, A. C., Canto, J., Calvet N.,
%Rodriguez, L.F., \& Torrelles, J.M., 1993b, \aap, 276, 539 
%\bibitem[Raga et al.(2004)]{Raga2004} Raga, A.~C., Noriega-Crespo, A., 
%Gonz{\' a}lez, R.~F., \& Vel{\' a}zquez, P.~F.\ 2004, \apjs, 154, 346 
%\bibitem[Raga, Cabrit \& Canto(1995)]{Raga1995} Raga, A. C., 
%Cabrit, S. \& Canto, J. 1995, \mnras, 273, 422
%\bibitem[Ray et al.(2007)]{Ray2007} Ray, T., Dougados, C., 
%Bacciotti, F., Eisl{\"o}ffel, J., 
%\& Chrysostomou, A.\ 2007, Protostars and Planets V, 231 
%\bibitem[Rawlings, Taylor, \& Williams(2000)]{Rawlings2000} 
%Rawlings, J.~M.~C., Taylor, S.~D., \& Williams, D.~A.\ 2000, \mnras, 313, 
%461
%\bibitem[Rawlings et al.(2004)]{Rawlings2004} Rawlings, J.~M.~C., 
%Redman, M.~P., Keto, E., \& Williams, D.~A.\ 2004, \mnras, 351, 1054 
%\bibitem[Rawlings et al.(2013)]{Rawlings2013} Rawlings, J.~M.~C., 
%Redman, M.~P., \& Carolan, P.~B.\ 2013, \mnras, 435, 289 
%\bibitem[Reipurth \& Olberg(1991)]{Reipurth1991} Reipurth, B. \& 
%Olberg, M.  1991, \aap, 246, 535 
%\bibitem[Reipurth \& Cernicharo(1995)]{Reipurth1995} Reipurth, B. \& 
% Cernicharo, J. 1995, Revista Mexicana de Astronomia y Astrofisica 
% Conference Series, 1, 43  
%\bibitem[Reipurth, Raga, \& Heathcote(1992)]{Reipurth1992} Reipurth, 
%  B., Raga, A. C. \& Heathcote, S.  1992, \apj, 392, 145 
%\bibitem[Reipurth Bally \& Devine(1997a)]{Reipurth1997a} Reipurth, B., 
%Bally, J.  \& Devine, D.  1997a, \aj, 114, 2708 
%\bibitem[Reipurth et al.(1993)]{Reipurth1993}Reipurth, B., Chini, R., Krugel, E., 
%  Kreysa, E., Sievers., A. 1993 A\&A, 273, 221
%\bibitem[Reipurth et al.(1997b)]{Reipurth1997b} Reipurth, B., Hartigan, 
%  P. , Heathcote, S. , Morse, J. A. \& Bally, J.  1997b, \aj, 114, 757 
%\bibitem[Reipurth et al.(1999)]{Reipurth1999} Reipurth, B., Yu, K. , 
%Rodriguez, L. F., Heathcote, S.  \& Bally, J.  1999, \aap, 352, L83 
%\bibitem[Reipurth et al.(2002)]{Reipurth2002} Reipurth, B., 
%Heathcote, S., Morse, J., Hartigan, P., \& Bally, J.\ 2002, \aj, 123, 362 
%\bibitem[Richer et al.(2000)]{Richer2000}Richer, J. Shepherd, D., 
%  Cabrit, S., Bachiller, R., \& Churchwell, E. 2000, in Protostars
%and Planets IV, ed. V. Mannings, A. P. Boss \& S. S. Russell
%  (Tucson: University of Arizona Press), in press 
%\bibitem[Rodriguez \& Reipurth(1994)]{Rodriguez1994} Rodriguez, L. F. 
% \& Reipurth, B. 1994, \aap, 281, 882 
%\bibitem[Rodriguez et al.(1998)]{Rodriguez1998} Rodriguez, L. F., 
% Reipurth, B., Raga, A. C. \& Canto, J. 1998, Revista Mexicana de Astronomia 
% y Astrofisica, 34, 69 
%\bibitem[Rodriguez \& Reipurth(1998)]{Rodriguez1998} Rodriguez, L. F., 
% \& Reipurth, B. 1998, Revista Mexicana de Astronomia 
% y Astrofisica, 34, 13 
%\bibitem[Sakai et al.(2014)]{Sakai2014} Sakai, N., Sakai, T., 
%Hirota, T., et al.\ 2014, \nat, 507, 78 
%\bibitem[Sandstrom et al.(2007)]{Sandstrom2007} Sandstrom, K.~M., 
%Peek, J.~E.~G., Bower, G.~C., Bolatto, A.~D., 
%\& Plambeck, R.~L.\ 2007, \apj, 667, 1161 
%\bibitem[Schilke et al.(1997)]{Schilke1997} Schilke, P., Walmsley, 
%C.~M., Pineau des Forets, G., \& Flower, D.~R.\ 1997, \aap, 321, 293 
%\bibitem[Shang et al.(1998)]{Shang1998} Shang, H. , Shu, 
%  F. H. \& Glassgold, A. E. 1998, \apjl, 493, L91 
%\bibitem[Shang et al.(2004)]{Shang2004} Shang, H., Lizano, S., 
% Glassgold, A., \& Shu, F.\ 2004, \apjl, 612, L69 
%\bibitem[Shang et al.(2006)]{Shang2006} Shang, H., Allen, A., Li, 
%Z.-Y., Liu, C.-F., Chou, M.-Y., \& Anderson, J.\ 2006, \apj, 649, 845 
%\bibitem[Shang et al.(2007)]{Shang2007} Shang, H., Li, Z.-Y., 
%\& Hirano, N.\ 2007, Protostars and Planets V, 261 
%\bibitem[Shirley et al.(2000)]{Shirley2000} 
% Shirley, Y.~L., Evans, N.~J., Rawlings, J.~M.~C., \& Gregersen, E.~M.\ 
% 2000, ApJS, 131, 249 
%\bibitem[Shu(1977)]{Shu1977} Shu, F.~H.\ 1977, \apj, 214, 488 
%\bibitem[Shu, Adams, \& Lizano(1987)]{Shu1987} Shu, F.~H., 
% Adams, F.~C., \& Lizano, S.\ 1987, \araa, 25, 23 
%\bibitem[Shu et al.(1991)]{Shu1991} Shu, F. H., 
%  Ruden, S. P., Lada, C. J. \& Lizano, S.  1991, \apjl, 370, L31 
%\bibitem[Shu et al.(1994)]{Shu1994} Shu, F., Najita, J., 
%Ostriker, E., et al.\ 1994, \apj, 429, 781 
%\bibitem[Shu et al.(1995) 1995]{Shu1995} Shu, F. 
%  H., Najita, J. , Ostriker, E. C. \& Shang, H.  1995, \apjl, 455, L155  
%\bibitem[Smith, Suttner, \& Yorke(1997)]{Smith1997} Smith, M. D., 
%  Suttner, G. \& Yorke, H. W. 1997, \aap, 323, 223 
%\bibitem[Smith \& Rosen(2007)]{Smith2007} Smith, M.~D., \& Rosen, 
%A.\ 2007, \mnras, 378, 691 
%\bibitem[Steer, Dewdney, \& Ito(1984)]{Steer1984} Steer, D.~G., 
% Dewdney, P.~E., \& Ito, M.~R.\ 1984, \aap, 137, 159 
%\bibitem[Stone \& Norman(1994)]{Stone1994} Stone, J. M. \& Norman, 
%M. L. 1994, \apj, 420, 237 
%\bibitem[Suttner et al.(1997)]{Suttner1997} 
%  Suttner, G., Smith, M. D., Yorke, H. W. \& Zinnecker, H. 1997, \aap, 318, 
%  595 
%\bibitem[Tafalla et al.(2002)]{Tafalla2002} Tafalla, M., Myers, 
%P.~C., Caselli, P., Walmsley, C.~M., \& Comito, C.\ 2002, \apj, 569, 815
%\bibitem[Takakuwa et al.(2003)]{Takakuwa2003} Takakuwa, S., 
%Kamazaki, T., Saito, M., \& Hirano, N.\ 2003, \apj, 584, 818
%\bibitem[Takami et al.(2006)]{Takami2006} Takami, M., Takakuwa, 
%S., Momose, M., Hayashi, M., Davis, C., Pyo, T.-S., Nishikawa, T., \& 
%Kohno, K.\ 2006, PASJ, ??
%\bibitem[Taylor 
%\& Raga(1995)]{Taylor1995} Taylor, S.~D., \& Raga, A.~C.\ 1995, \aap, 296, 823 
%\bibitem[Terebey et al.(1993)]{Terebey1993} Terebey, S., 
%  Chandler, C. J. \& Andre, P. 1993, \apj, 414, 759 
%\bibitem[Todo et al.(1993)]{Todo1993} Todo, Y., Uchida, Y., 
%Sato, T., \& Rosner, R.\ 1993, \apj, 403, 164 
%\bibitem[Ulrich(1976)]{Ulrich1976} Ulrich, R.~K.\ 1976, \apj, 210, 
%377 
%\bibitem[Umemoto et al.(1991)]{Umemoto1991} Umemoto, T., Hirano, 
%  N., Kameya, O., Fukui, Y., Kuno, N. \& Takakubo, K. 1991, \apj, 377, 510
%\bibitem[Velusamy \& Langer(1998)]{Velusamy1998} Velusamy, T. \& 
%  Langer, W. D. 1998, \nat, 392, 685 
%\bibitem[Viti, Natarajan, \& Williams(2002)]{Viti2002} Viti, S., 
% Natarajan, S., \& Williams, D.~A.\ 2002, \mnras, 336, 797 
%\bibitem[\Volker{} et al.(1999)]{Volker1999}
%  \Volker, R. , Smith, M. D., Suttner, G.  \& Yorke, H. W. 1999, 
%  \aap, 343, 953 
%\bibitem[Ward-Thompson, Motte, \& Andre(1999)]{Ward-Thompson1999} 
% Ward-Thompson, D., Motte, F., \& Andre, P.\ 1999, \mnras, 305, 143 
%\bibitem[Weintraub et al.(1994)]{Weintraub1994} 
%  Weintraub, D. A., Tegler, S. C., Kastner, J. H. \& Rettig, T.  1994, \apj, 
%423, 674 
%\bibitem[Wootten et al.(2002)]{Wootten2002} 
% Wootten, A., Mangum, J.~G., Wiseman, J., \& Fuller, G.~A.\ 2002, Bulletin 
% of the American Astronomical Society, 34, 1134 
%\bibitem[Wouterloot \& Walmsley(1986)]{Wouterloot1986} Wouterloot, J. 
%  G. A. \& Walmsley, C. M. 1986, \aap, 168, 237 
%\bibitem[Wu, Huang \& He(1996)]{Wu1996} Wu, Y., Huang, M. \& He, 
%  J. 1996, \aaps, 115, 283 
%\bibitem[Xu et al.(2000)]{Xu2000} Xu, J., Hardee, P.~E., \& 
%Stone, J.~M.\ 2000, \apj, 543, 161 
%\bibitem[Yang et al.(1997)]{Yang1997} Yang, J., Ohashi, N. , Yan, 
% J. , Liu, C. , Kaifu, N.  \& Kimura, H.  1997, \apj, 475, 683 
\end{thebibliography}
\end{document}